\documentclass[aps,preprint,prd,nofootinbib,epsfig]{revtex4}
\pdfoutput=1 
\usepackage{pifont}
\usepackage{color}
\usepackage{amssymb}
\usepackage{hyperref}
\usepackage{amsmath}
\usepackage{graphicx}
\usepackage{fancyhdr}
\newcommand{\be}{\begin{equation}}
\newcommand{\ee}{\end{equation}}
\newcommand{\bea}{\begin{eqnarray}}
\newcommand{\eea}{\end{eqnarray}}
\newcommand{\bes}{\begin{subequations}}
\newcommand{\ees}{\end{subequations}}

\newcommand{\bc}{\begin{center}}
\newcommand{\ec}{\end{center}}
\begin{document}
%%%%%
\title{  On the Higgs-like boson in the  Minimal Supersymmetric 3-3-1 Model}
%%%%%
\author{J. G. Ferreira Jr$^a$, C. A. de S. Pires$^a$, P. S. Rodrigues da Silva$^a$, Clarissa Siqueira$^{a,b}$}
\affiliation{$^a$Departamento de F\'{\i}sica, Universidade Federal da Para\'\i ba, Caixa Postal 5008, 58051-970,
Jo\~ao Pessoa, PB, Brazil}
\affiliation{$^b$Max-Planck-Institut f\"ur Kernphysik, Saupfercheckweg 1, 69117 Heidelberg, Germany}
% e-mail addresses: one for each author, in the same order as the authors
%\emailAdd{cpires@fisica.ufpb.br}
%\emailAdd{psilva@fisica.ufpb.br}
%\emailAdd{antonio$\_$santos@fisica.ufpb.br}
%emailAdd{clarissa@fisica.ufpb.br}

\date{\today}

\begin{abstract}
It is imperative that any  proposal of new physics possesses a Higgs-like boson with 125 GeV of mass and couplings with the standard particles that recover the  branching ratios and signal strengths as measured by CMS and ATLAS. We address this issue within the supersymmetric version of the minimal 3-3-1 model. For this we develop the Higgs potential  with focus on the lightest Higgs provided by the model. Our proposal is to  verify if it recovers the properties of the standard Higgs.  With respect to its mass, we calculate it up to one loop level by taking into account all contributions provided by the model. In regard to  its couplings, we restrict our investigation to  couplings of the Higgs-like boson with the standard particles, only. We then  calculate the dominant branching ratios and the respective signal strengths and  confront our results with the recent measurements of CMS  and ATLAS.  As distinctive aspects, we remark that our  Higgs-like boson  intermediates flavor changing neutral processes and then argue that its signature is the decay $t \rightarrow h+c$. We calculate its branching ratio and compare it with current bounds. We also show that   the potential is stable for the region of parameter space employed in our calculations.
\end{abstract}

\maketitle

\section{Introduction}
The discovery of a scalar state by the ATLAS and CMS  at CERN  \cite{Aad:2012tfa,Chatrchyan:2012xdj} compatible with the standard model Higgs boson was a milestone in particle physics since it was the missing part of the theory. Nowadays we not only know the Higgs mass with precision, $M_h = 125.09\pm0.21 (stat.) \pm 0.11 (syst.)$ GeV, but also know that its signal strengths are found consistent with the standard model (SM) predictions\cite{Khachatryan:2016vau}. However, the present experimental uncertainties allow us to interpret  such observed state as belonging to a more complex theory that encompasses the standard one.  

Supersymmetry is the most popular proposal of new physics and all supersymmetric models predict at least one light Higgs-like boson. For example, the minimal supersymmetric standard model (MSSM) provides a tree level contribution to the mass of the Higgs-like boson that does not surpass  the neutral gauge boson $Z^0$ mass. Consequentely the amount of radiative corrections required to complete 125 GeV demands  stops with mass in a region of values that threaten the naturalness principle\cite{Carena:2002es}. In view of this it turns out imperative to examine the Higgs sector of any supersymmetric phenomenological model with the aim of investigating  the lightest Higgs concerning its mass and  couplings with the standard particles.

In this work we develop the scalar sector of the supersymmetric version of the minimal $SU(3)_C \times SU(3)_L \times U(1)_N$ (331) model. The motivations to study 331 models are various as, for example, explanation of family replication\cite{Pisano:1991ee}\cite{Frampton:1992wt}, electric charge quantization\cite{deSousaPires:1998jc}\cite{deSousaPires:1999ca}, strong CP-problem\cite{Pal:1994ba}\cite{Dias:2003zt}\cite{Dias:2003iq}, incorporation of inflation\cite{Ferreira:2016uao}  and so forth. As we show in this paper, its SUSY versions, besides solving the hierarchy problem, provides a tree level contribution to the lightest Higgs boson that may surpass 100 GeV. This is nice in what concerns the naturalness principle since now 125 GeV Higgs is compatible with stop with mass below the TeV scale.

The minimal supersymmetric 331 model was firstly developed in \cite{Duong:1993zn}. For subsequent works, see \cite{Montero:2000ng}\cite{Huong:2012pg}\cite{Ferreira:2013nla}. As the Higgs sector of the minimal 331 model\cite{Pisano:1991ee}\cite{Frampton:1992wt} requires at least  three Higgs triplets to generate correctly the masses of all massive particles and be phenomenologically viable\cite{Montero:2001tq}\cite{Queiroz:2010rj}\cite{Dong:2014bha}, then its supersymmetric version, what we call the minimal SUSY331 model,  must involve six Higgs triplets as required to cancel anomalies. On assuming  that the two triplets $\chi$ and $\chi^{\prime}$ decouple from the other four $\eta$, $\rho$, $\eta^{\prime}$ and $\rho^{\prime}$, and that  $\eta$ and $\rho$ are inert, then after  spontaneous breaking of the $331$ symmetry to the standard one, we obtain the potential for two Higgs doublets modified by cubic invariant terms. On developing such potential, we obtain the following approximate expression for the lightest scalar, the  Higgs-like boson candidate, of the model
\be
M^2_h\leq 4M^2_Z \sin^2 (\theta_w)(1 +\cos^2(\beta))^2,
\label{ma}
\ee
where $M_Z$ is the mass of the standard neutral gauge boson, $\theta_w$ is the electroweak mixing angle and the angle $\beta$ is such that $\tan \beta= \frac{v_{\eta^{\prime}}}{v_{\rho^{\prime}}}$.  According to this expression the tree level contribution to the  lightest Higgs may attain  174 GeV for $\cos(\beta)=1$.This is so because  the superpotential naturally contains cubic invariant terms that act like in the NMSSM\cite{Duong:1993zn}.   In view of this, it is compulsory to review the Higgs sector of this model in the light of the recent discoveries of the LHC, and verifying how realistic the expression above is and define precisely the profile of the lightest Higgs in the minimal SUSY331 model.  

Our contribution to this discussion restricts to obtain the mass matrix that contains the neutral Higgs,  diagonalize it in the most general way and obtain the eigenvalue and the  eigenvector that corresponds to the lightest Higgs. We investigate the behavior of its mass by taking into account one loop-corrections. We also obtain the branching ratios for the dominant processes  and calculate the respective signal strengths. It is notorious that our Higgs-like boson intermediates flavor changing processes.  We discuss such processes but focus on the dominant ones  that are $t \rightarrow h+c$ and $t \rightarrow h +u$.  Finally, we guarantee that the region of parameter space employed in our calculations leads to a stable potential. 

The paper is divided in the following way: In  Sec.~\ref{sec1} we present the main ingredients of the model. Next, in Sec.~\ref{sec2}, we develop the Higgs sector with focus on the lightest Higgs where we calculate its mass up to one-loop level. In Sec.~\ref{sec3} we calculate its branching ratios and the respective signal strengths. In Sec.~\ref{sec4} we calculate the flavor changing processes. In Sec.~\ref{sec5}  we  address the stability of the potential and,  finally, we conclude in Sec.~\ref{sec6}.

\section{The main ingredients of the model}
\label{sec1}
%%%%%%

In this section we present the core of the minimal SUSY331 model relevant for what follows. In regard to the leptonic sector, the superfields of each generation are arranged in triplet of superfields according to the following transformation by the 3-3-1 symmetry
%%%%
\begin{eqnarray}
\hat L_{a} = \left (
\begin{array}{c}
\hat \nu_{a} \\
\hat e_{a} \\
\hat e_{a}^c
\end{array}
\right )_L\sim(1\,,\,3\,,\,0)\,,
\label{Lcontent}
\end{eqnarray}
where $a=1,2,3$ represents the family index for the usual three generations of leptons. 

In the Hadronic sector, the superfields of the  third generation comes in the triplet representation and the superfields of the other two come in anti-triplet representations of $SU(3)_L$, as a requirement for anomaly cancellation. They are given by
\begin{eqnarray}
&& \hat{Q}_{\alpha_L}=\left(\begin{array}{c}
                 \hat{d}_{\alpha}  \\
                 \hat{u}_{\alpha}  \\
                 \hat{j}^{\prime}_{\alpha}
                 \end{array}\right)_L \sim (3,3^*,-\frac{1}{3}) , \nonumber \\
                 && \hat u_{\alpha_L}^{\,c} \sim(3^*,1,-\frac{2}{3})\,,\,\,\, \hat d_{\alpha_L}^{\,c} \sim(3^*,1,1/3)\, ,\,\,\hat j^{\prime{\,c}}_{\alpha_L} \sim(3^*,1,\frac{4}{3})\,,\,\,\, \nonumber \\
&& \hat{Q}_{3 L}=\left(\begin{array}{c}
                 \hat{u}_3  \\
                 \hat{d}_3  \\
                 \hat{j}^{\prime}_3 
                \end{array}\right)_L \sim (3,3,\frac{2}{3}), \nonumber \\
&& \hat u_{3_L}^{\,c} \sim(3^*,1,-\frac{2}{3})\,,\,\,\, \hat d_{3_L}^{\,c} \sim(3^*,1,1/3)\, ,\,\,\hat j_{3L}^{\,c}\sim(3^*,1,-\frac{5}{3}),
\end{eqnarray}
where $\alpha = 1,2$.

The scalar sector of the 3-3-1  model, responsible for the spontaneously broken gauge symmetry, is composed by three scalar triplets.  In the supersymmetric version, anomaly cancellation requires we  double these fields. Thus, the scalar sector of the minimal SUSY331 is composed by the following superfields

\begin{eqnarray}
 \label{part_transf1}
 \hat{\eta}=\left(\begin{array}{c}
                 \hat{\eta}  \\
                 \hat{\eta}_{1}^{-}  \\
                 \hat{\eta}_{2}^{+} \\
                 
\end{array}\right) , 
 \hat{\chi}=\left(\begin{array}{c}
                 \hat{\chi}^{-}  \\
                 \hat{\chi}^{--}  \\
                 \hat{\chi} \\
                 
\end{array}\right), 
 \hat{\rho}=\left(\begin{array}{c}
                 \hat{\rho}^{+}  \\
                 \hat{\rho}  \\
                 \hat{\rho}^{++} \\
                 
\end{array}\right),
\label{triplets}
\end{eqnarray}
where $\hat{\eta} \sim (1,3,0)\,, \hat{\chi}\sim (1,3,-1)\,,\,\,\,$  $\hat{\rho}\sim (1,3,1)$, and
\begin{eqnarray}
 \label{part_transf2}
 \hat{\eta}^{\prime}=\left(\begin{array}{c}
                 \hat{\eta}^{\prime}  \\
                 \hat{\eta}_{1}^{\prime +}  \\
                 \hat{\eta}_2^{\prime -} \\
                 
\end{array}\right) , 
 \hat{\chi^{\prime}}=\left(\begin{array}{c}
                 \hat{\chi}^{\prime +}  \\
                 \hat{\chi}^{\prime ++}  \\
                 \hat{\chi}^{\prime} \\
                 
\end{array}\right), 
 \hat{\rho}^{\prime}=\left(\begin{array}{c}
                 \hat{\rho}^{\prime -}\!\!\! \\
                 \hat{\rho}^{\prime} \\
                 \hat{\rho}^{\prime --}\!\!\! \\
                 
\end{array}\right),
\label{antitriplets}
\end{eqnarray}
where $\hat{\eta}^{\prime} \sim (1,3^*,0)\,, \hat{\chi}^{\prime} \sim (1,3^*,1)\,,\,\,\,$  $\hat{\rho}^{\prime} \sim (1,3^*,-1)$.  

We assume that the neutral scalars  $\eta$, $\eta^{\prime}$,  $\rho$, $\rho^{\prime}$,  $\chi$ and $\chi^{\prime}$ develop nonzero VEV according to
\begin{eqnarray}
 \langle\eta\rangle=\frac{v_{\eta}}{\sqrt2},\,\,
\langle\eta^{\prime}\rangle=\frac{v_{\eta^{\prime}}}{\sqrt2},\,\,
 \langle \rho \rangle=\frac{v_{\rho}}{\sqrt2},\,\,
  \langle \rho^{\prime} \rangle=\frac{v_{\rho^{\prime}}}{\sqrt2},\,\,\langle\chi\rangle=\frac{v_{\chi}}{\sqrt2},\,\,
\langle\chi^{\prime}\rangle=\frac{v_{\chi^{\prime}}}{\sqrt2}.
\label{vevs}
\end{eqnarray}
These VEVs lead to the following gauge symmetry breaking pattern
\begin{equation}
SU(3)_C \otimes SU(3)_L \otimes U(1)_X  \stackrel{v_{\chi}, v_{\chi^{\prime}}}{ \Longrightarrow } SU(3)_C \otimes SU(2)_L \otimes U(1)_Y \stackrel{v_{\eta}, v_{\eta^{\prime}},v_{\rho},v_{\rho^{\prime}} }{ \Longrightarrow } SU(3)_C \otimes U(1)_{QED}.
\label{symmetrygreakingpattern}
\end{equation}

With the breaking of the gauge symmetry by this set of VEVs all the massive particles, including the supersymmetric ones, receive mass.  What matters for us here are the scalar and gauge boson  masses. Concerning the gauge bosons, they are composed by the standard gauge boson, $\gamma$, $ Z^0$ and $W^{\pm}$, one new neutral massive gauge bosons $Z^{\prime}$, two doubly charged gauge bosons $U^{\pm\pm}$, and two simply charged gauge bosons $V^{\pm}$. The charged gauge bosons gain the following  mass expressions
\begin{eqnarray}
 M_{W^{\pm}}&=&\frac{g^2}{4}\left(v_{\eta}^2+v_{{\eta^\prime}}^2+v_{\rho}^2+v_{{\rho^\prime}}^2 \right) \nonumber \\
  M_{V^{\pm}}&=&\frac{g^2}{4}\left(v_{\eta}^2+v_{{\eta^\prime}}^2+v_{\chi}^2+v_{{\chi^\prime}}^2 \right) \nonumber \\
   M_{U^{\pm\pm}}&=&\frac{g^2}{4}\left(v_{\rho}^2+v_{{\rho^\prime}}^2+v_{\chi}^2+v_{{\chi^\prime}}^2 \right),
\end{eqnarray}
while the neutral gauge bosons have masses
\begin{eqnarray}
 M_{Z} &=& g^2 \frac{(1+4 t^2)}{(4+12 t^2)}\left(v_{\eta}^2+v_{{\eta^\prime}}^2+v_{\rho}^2+v_{{\rho^\prime}}^2 \right) \nonumber \\
 M_{Z^\prime} &=& \frac{1}{3}g^2(3 t^2 +1)(v_{\chi}^2+v_{{\chi^\prime}}^2),
\end{eqnarray}
where  $t=g_N/g$,  $v_\rho^2+v^2_{ \rho^{\prime}}+v_{\eta}^2+v_{\eta^{\prime}}^{ 2}=v^2_{ew}$ and $v_{\chi}^2+v_{\chi^{\prime}}^2\equiv v^2_{331}$ with $v_{331}$ lying in the TeV scale.

After all this, we are ready to build up  the superpotential of the model. The superpotential that respects the gauge symmetry and R-parity is composed by the following terms
 \begin{eqnarray}
 W&=& \mu_{\eta} \hat{\eta}\hat{\eta}^\prime + \mu_{\rho} \hat{\rho}\hat{\rho}^\prime + \mu_{\chi} \hat{\chi}\hat{\chi}^\prime +
 f_1 \hat{\rho}\hat{\chi}\hat{\eta}+
 f_1^\prime \hat{\rho}^\prime\hat{\chi}^\prime\hat{\eta}^\prime + \sum_i \left(\frac{\kappa_{1ij}}{\Lambda} \hat{L}_j\hat{\rho}^{\prime}\hat{\chi}^\prime\hat{L}_i\right) \nonumber \\ &+&
 \sum_{i,\alpha}\left(
 \kappa_{2 i \alpha} \hat{Q}_\alpha \hat{\eta} \hat{d}^c_i + 
 \kappa_{3 i \alpha} \hat{Q}_\alpha \hat{\rho} \hat{u}^c_i \right) + \sum_{\alpha,\beta} \left(
 \kappa_{4 \alpha \beta} \hat{Q}_\alpha \hat{\chi} \hat{j}^c_\beta
 \right) \nonumber\\
  &+&
 \sum_i \left(\kappa_{5 i} \hat{Q}_3\hat{\eta}^\prime \hat{u}_i^c +\kappa_{6 i} \hat{Q}_3\hat{\rho}^\prime \hat{d}_i^c \right) + \kappa_{6} \hat{Q}_3 \hat{\chi}^{\prime} \hat{j}_3^c ,
 \label{superpotential}
\end{eqnarray}
where $\alpha, \beta=1,2$ and $i=1,2,3$. We could be economical and resort to a $Z_3$ symmetry so as to avoid the bilinear terms in the superpotential above. However, in order to be general enough, we do not follow this path  here.

We call the attention to the fact that the last term in the first line of the superpotential is an effective 5-D operator. It will generate the masses of the charged leptons. This point has been well developed in many previous papers, but it is appropriate to recall it here. The  highest energy scale where the model is found to be perturbatively reliable is about $\Lambda=4-5 $TeV. Hence, effective operators may be required to generate corrections  to  the mass of some charged fermions. We make use of this here with the aim  of avoiding the addition  of scalars sextets to the model. 

Up to this point the masses of the standard particles are equal to the masses of their superpartners. 
As usual, in phenomenological supersymmetric models SUSY must be broken so as to provide a reasonable shift between ordinary particles and their supersymmetric partners. In this work we assume that SUSY is broken explicitly through the following set of soft breaking terms that are invariant under the symmetries  assumed here
\begin{eqnarray}
\mathcal{L}_{soft} &=& -\frac{1}{2}\left( m_{\lambda_c} \sum_a \left(\lambda_c^a \lambda_c^a \right) + m_\lambda \sum_a \left(\lambda^a \lambda^a \right) + m_\lambda^\prime \lambda \lambda + h.c. \right) \nonumber \\ 
&-& m_{L}^2 \tilde{L}^\dagger \tilde{L} - m_{Q_\alpha}^2 \tilde{Q_\alpha}^\dagger \tilde{Q_\alpha} + \sum_i \left(\tilde{u}^\dagger_i m_{u_i}^2 \tilde{u}_i - \tilde{d}^\dagger_i m_{d_i}^2 \tilde{d}_i \right)
- m_J^2 \tilde{j}_3^\dagger \tilde{j}_3 \nonumber \\ 
&-& \sum_\beta \tilde{j_\beta} m_{j \beta}^{2} \tilde{j}_\beta - \sum_\alpha m_{Q_3}^2 \tilde{Q}_3^\dagger \tilde{Q}_3 - m_\eta^2 \eta^\dagger \eta - m_\rho^2 \rho^\dagger \rho  - m_\chi^2 \chi^\dagger \chi \nonumber \\ 
&-&  m_\eta^{\prime \, 2} \eta^{\prime \dagger} \eta^{\prime} - m_\rho^{\prime \, 2} \rho^{\prime \dagger} \rho^{\prime}  - m_\chi^{\prime \, 2} \chi^{\prime \dagger} \chi^{\prime \dagger}+
k_1 \rho \chi \eta + k_2 \rho^\prime \chi^\prime \eta^\prime \nonumber \\
&+& b_{\rho} \rho^\prime \rho + b_{\eta} \eta^\prime \eta + b_{\chi} \chi^\prime \chi + \sum_\alpha \tilde{Q}_\alpha [ \sum_i (\omega_{1\alpha i} \eta \tilde{d}^c_i + \omega_{2\alpha i} \rho \tilde{u}^c_i )  + \sum_\beta \omega_{3 \alpha \beta} \chi \tilde{j}_{\beta}^c + h.c. ]\nonumber \\
&+&  \sum_i \tilde{Q}_3(\xi_{1i}\eta^\prime \tilde{u}^c_i + \xi_{2i}\rho^\prime \tilde{d}^c_i + \xi_{3}\chi^\prime \tilde{j}_3^c). \nonumber
\label{lsoft}
\end{eqnarray}
where $\alpha, \beta=1,2$ and $i=1,2,3$. $\lambda_c^a$ are the gluinos, $\lambda^a$ are gauginos associated to $SU(3)_L$ (in both cases $a=1,...,8$ is the gauge group index) and $\lambda$ is the gaugino associated to $U(1)_N$. The  scalar supersymmetric partners of fermion fields, $f$, are denoted by $\tilde{f}$, while the remaining fields are self-evident. For simplicity, we will take $m_{L}^2=m_{Q_\alpha}^2=m_{Q_3}^2=m_{SUSY_L}^2$, $m_{u_i}^2=m_{d_i}^2= m_{j \beta}^2=m_{SUSY_R}^2$ and $m_J^2=10 \times m_{SUSY_R}^2$.

As it is usual in SUSY models, which involves  a large number of free parameters,  simplifications are necessary in order to easily get a better view of the big picture. As simplification we use the following parametrization
\begin{eqnarray}
 b_{\rho,\eta,\chi}&=& B_0 \times \mu_{\rho,\eta,\chi}\,\,; \,\,\,\,\,\,  \omega,\xi = A_0 \times \kappa ; \,\,\,\,\,M_{s} = (m_{\tilde{t}_{1}}m_{\tilde{t}_{2}})^{1/2} , \nonumber\\
v_{331}&=&\sqrt{v_\chi^2 + v^2_{\chi^{\prime }}} ; \,\,\,\,\,\,\tan{\beta}= \frac{\sqrt{v_{\eta}^2 + v^2_{{\eta}^{\prime }}}}{\sqrt{v_{\rho}^2 + v^2_{{\rho}^{\prime }}}} , \nonumber \\
 \tan{\beta_\eta}&=&\frac{v_{\eta^{\prime}}}{v_\eta} ; \,\,
\tan{\beta_\rho}=\frac{v_{\rho^{\prime}}}{v_\rho} ; \,\,
\tan{\beta_\chi}=\frac{v_{\chi^{\prime}}}{v_\chi} , \nonumber \\
X_{t}& =& \xi_{13} + \mu_\eta \cot{\beta_\eta}+f_1^\prime v_{331}\cot{\beta} \frac{\sin{\beta_\rho}\sin{\beta_\chi}}{\sin{\beta_\eta}}, 
\label{paramete}
\end{eqnarray}
where $\kappa$'s are the corresponding Yukawa couplings,  $M_s$ is the SUSY scale,  $m_{\tilde{t}}$ is the mass of the stop and $X_{t}$ is the equivalent of the soft trilinear coupling of the stops. We also assume that  all soft left and right masses, $m_{SUSY_L}$ and $m_{SUSY_R}$, are equals. With all this we are ready to further explore the Higgs physics.
\section{ Higgs physics in the minimal SUSY331 model}
\label{sec2}
In supersymmetric models the Higgs potential receives contributions from three different sources, F-term, D-term and the soft SUSY breaking terms,  that adds up to form the  scalar potential $V= V_{F}+V_{D}+V_{soft}$.  These contributions are, respectively
\bea
 V_F &=& \mu_\eta^2 |\eta|^2 + \mu_\eta^2 |\eta^\prime|^2 + \mu_\rho^2 |\rho|^2 + \mu_\rho^2 |\rho^\prime|^2 + \mu_\chi^2 |\chi|^2 + \mu_\chi^2 |\chi^\prime|^2 \nonumber \\ 
&+& f_1^2 \left(|\eta|^2 |\rho|^2 - |\eta \cdot \rho|^2 \right) + f_1^2 \left(|\eta|^2 |\chi|^2 - |\eta \cdot \chi|^2 \right) + f_1^2 \left(|\chi|^2 |\rho|^2 - |\chi \cdot \rho|^2 \right) \nonumber \\
 &+& f_1^{\prime 2} \left(|\eta^\prime|^2 |\rho^\prime|^2 - | \eta^{\prime} \cdot \rho^\prime
| ^2 \right) + f_{1}^{\prime 2} \left(|\eta^\prime|^2 |\chi^\prime|^2 - |\eta^\prime \cdot \chi^\prime |^2 \right) + f_1^{\prime 2} \left(|\chi^\prime|^2 |\rho^\prime|^2 - |\chi^\prime \cdot \rho^\prime|^2 \right) \nonumber \\
 &-&
 f_1 \epsilon_{ijk} (\mu_\eta \eta^{\prime \dagger}_i \rho_j \chi_k + \mu_\rho \eta_i \rho^{\prime \dagger}_j \chi_k + \mu_\chi \eta_i \rho_j \chi^{\prime \dagger}_k + \mathrm{h.c.} ) \nonumber \\
 &-& f_1^\prime \epsilon_{ijk} (\mu_\eta \eta_i \rho^{\prime \dagger}_j \chi^{\prime \dagger}_k + \mu_\rho \eta^{\prime \dagger}_i \rho_j \chi^{\prime \dagger}_k + \mu_\chi \eta^{\prime \dagger}_i \rho^{\prime \dagger}_j \chi_k + \mathrm{h.c.}), \nonumber \\ \\
V_{D} &=& \frac{g^{2}}{2}\sum_A\left( \rho^{{}_{\dagger}}t_{A}\rho - \rho^{{}_{\prime}{}_{\dagger}}t^{\ast}_{A}\rho^{{}_{\prime}}+\eta^{{}_{\dagger}}t_{A}\eta - \eta^{{}_{\prime}{}_{\dagger}}t^{\ast}_{A}\eta^{{}_{\prime}}  + \chi^{{}_{\dagger}}t_{A}\chi - \chi^{{}_{\prime}{}_{\dagger}}t^{\ast}_{A}\chi^{{}_{\prime}}\right)^{2} \nonumber \\ 
&+& \frac{g_{N}^{2}}{2}\left( \rho^{{}_{\dagger}}\rho - \rho^{{}_{\prime}{}_{\dagger}}\rho^{{}_{\prime}} - \chi^{{}_{\dagger}}\chi + \chi^{{}_{\prime}{}_{\dagger}}\chi^{{}_{\prime}}\right)^{2} , \nonumber \\ \\
V_{soft}&= &- m_\eta^2 \eta^\dagger \eta - m_\rho^2 \rho^\dagger \rho  - m_\chi^2 \chi^\dagger \chi  - m^2_{\eta^{\prime }} \eta^{\prime \dagger} \eta^{\prime} - m^2_{\rho^{\prime }} \rho^{\prime \dagger} \rho^{\prime}  - m^2_{\chi^{\prime }} \chi^{\prime \dagger} \chi^{\prime } \nonumber \\
&+& b_{\rho} \rho^\prime \rho + b_{\eta} \eta^\prime \eta + b_{\chi} \chi^\prime \chi +
k_1 \rho \chi \eta + k_2 \rho^\prime \chi^\prime \eta^\prime ,
\label{potential}
\eea
where $t_A$ are the Gell-Mann matrices.

The fields are assumed to be shifted in the usual way

\be
\rho^0,  \rho^{\prime 0}, \eta^0 , \eta^{\prime 0 } \chi^0 , \chi^{\prime 0} \rightarrow \frac{1}{\sqrt{2}} (v_{ \rho,\rho^{\prime },\eta,\eta^{\prime} \chi , \chi^{\prime}} + R_{ \rho,\rho^{\prime }, \eta, \eta^{\prime} \chi , \chi^{\prime}} + iI_{ \rho,\rho^{\prime }, \eta , \eta^{\prime}\chi , \chi^{\prime}}),
\label{scalarshift}
\ee
and the set of minimum conditions are given in {\bf Appendix A}.  That set of constraint  equations enable us to obtain the mass matrix of the scalars of the model.

In this work we focus exclusively in the lightest neutral scalar which is expected to recover the properties of the standard Higgs in what concerns its mass, branching ratios and signal strengths.  For this we first  obtain the $6 \times 6$ mass matrix, $M^2_{R_H}$,  associated to the CP-even scalars $R_\eta$, $R_{\eta^{\prime}}$,  $R_\rho$, $R_{\rho^{\prime}}$,  $R_\chi$ and $R_{\chi^{\prime}}$. It is a very complex, and not illuminating, mass matrix, then we do not show it here. Next, we diagonalize it by a rotating mixing matrix $U_R$ such that the diagonal mass matrix $M^2_{H_D} $ relates to the $M^2_{R_H}$ through the relation $M^2_{H_D} =U^{\dagger}_R M^2_{R_H} U_R$. The physical eigenstates relates to the symmetrical ones by  $H=U_R R$ where $H=\left( h_1,h_2,h_3,h_4,h_5,h_6 \right)^T$ and $R=\left( R_\eta, R_{\eta^{\prime}}, R_\rho, R_{\rho^{\prime}}, R_\chi, R_{\chi^{\prime}}\right)^T$. The lightest eigenstate, let us call it $h_i$,  must be identified as the Higgs-like boson. This means that it must have a mass of $125$ GeV and its eigenvector, $h_i=(U_R)_{ij}R_j$, must recover the standard  Higgs couplings.  In practical terms,  to know the eigenvector $h_i$ means to obtain the set of entries $(U_R)_{ij}$.  Such entries, when combined with the adequate couplings, must recover the existing  branching ratios and signal strengths of the Higgs. From now on we refer to the eigenvector associated to the Higgs-like boson as $h$.

 Due to the large number of  parameters involved in the diagonalisation of $M^2_{R_H}$, an analytical approach is unviable, and we turn it  completely numerical. In what follows we make use of powerful tools like the packages Sarah \cite{Staub:2008uz,Staub:2015kfa,Vicente:2015zba}, Spheno \cite{Porod:2003um,Porod:2011nf} and SSP \cite{Staub:2011dp}. Throughout this work we employ the following routine:  we implement the model in the Sarah and export it to the Spheno, where we make all the numerical calculations. The SSP helps in making the scan of the parameters. Hereafter  all numerical computatiions  done in this paper will follow this routine, including the diagonalisation of  this mass matrix.
%%%%%%%%%%%%%%%%%%%%%
\subsection{Tree Level Contribution}
%%%%%%%%%%%%%%%%%
Differently from the MSSM, where the tree level contribution to the Higgs mass  does not surpass $91$ GeV, in the minimal SUSY331 model things are a little different. The expression in Eq. (\ref{ma}) predicts that the lightest scalar, the Higgs-like boson,  may attain a mass  of $125$ GeV already at tree level. It is important  to note that the approximations assumed in Eq. (\ref{ma}) do not take into account the stability of the potential. Thus, it becomes mandatory, and challenging,  to clarify  if the general case,  which involves the diagonalisation of a $6 \times 6$ mass matrix, agrees with the prediction in Eq. (\ref{ma}).

\begin{figure}[!h]
 \includegraphics[width=8cm,height=6cm]{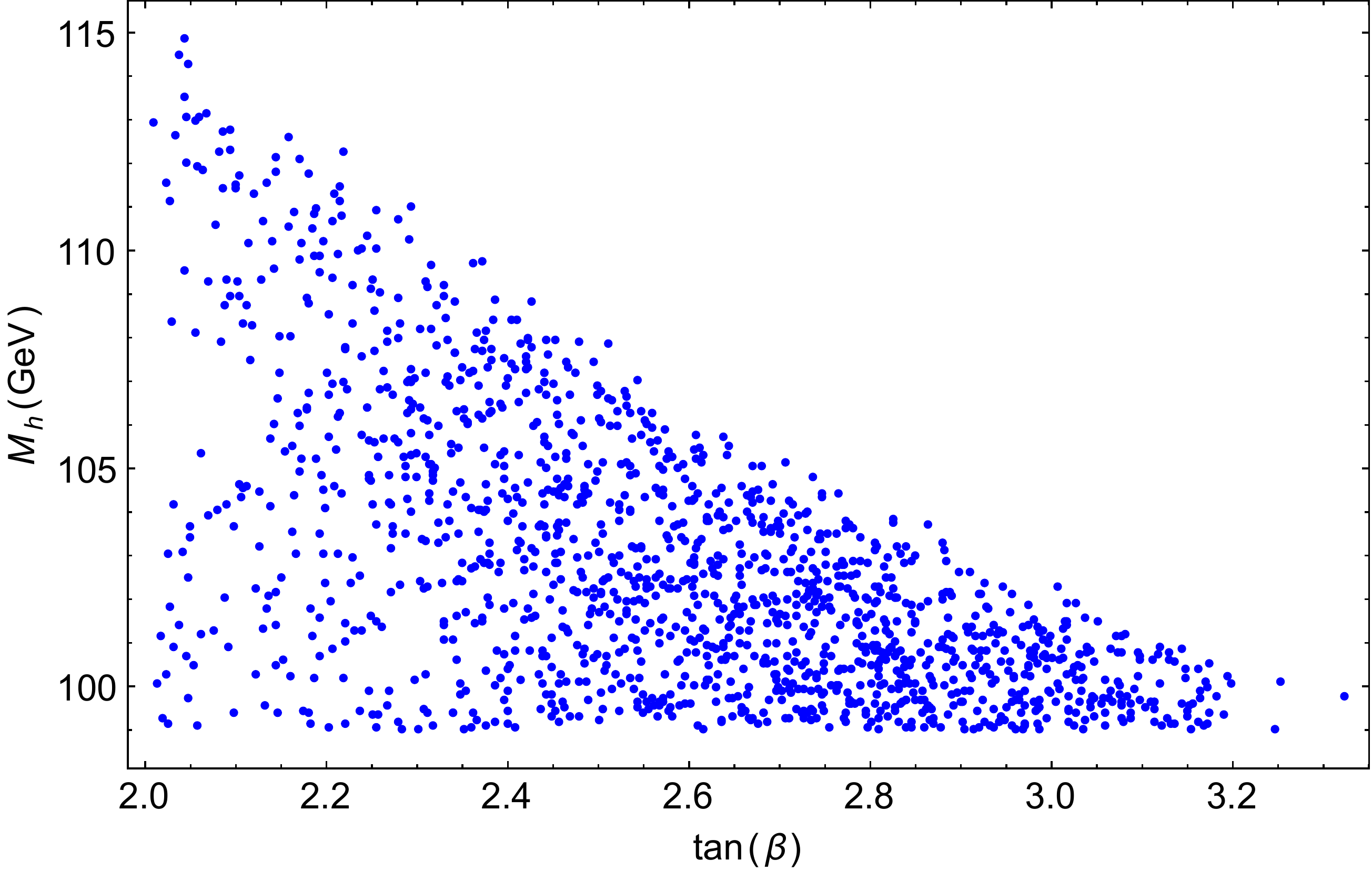}
 \caption{Tree level contribution to the Higgs-like boson mass as function of $\tan (\beta)$. }
 \label{tree}
\end{figure}

The  results  displayed in FIG. \ref{tree} is for the range of parameters 
\begin{eqnarray}
\centering
-0.03 \leq  f_1 \leq 0.2\,&,& 0.5 \leq f_1^{\prime} \leq 0.6\,,  \nonumber \\
-2000\,\mathrm{GeV} \leq k_1 \leq 2000\,\mathrm{GeV}\,&,&-1000\,\mathrm{GeV} \leq k_2 \leq 500 \, \mathrm{GeV}\,,\nonumber \\
-800\,\mathrm{GeV}  \leq \, \mu_{\eta} \,  \leq -500\,\mathrm{GeV}\,,-1800\,\mathrm{GeV}  \leq \, &\mu_{\rho}& \, \leq -1600\,\mathrm{GeV}\,,-1800\,\mathrm{GeV} \leq  \mu_{\chi}  \leq -1400\,\mathrm{GeV}\,, \nonumber \\ 
2 \leq  \tan{\beta}\leq 4,\,\,\, 8 \leq  (\tan{\beta_\eta}&,&\tan{\beta_\rho})\leq 10, \,\,\,0.95 \leq \tan{\beta_\chi} \leq 1.20 , \nonumber \\
 v_{\rho}^2+v_{\eta}^2+v_{\eta^{\prime}}^2+v_{\rho^{\prime}}^2&=&(246\,\mathrm{GeV})^2\,\,,\,\,A_0, B_0 = 1000\,\mathrm{GeV}.
\label{scan1}
\end{eqnarray}

Observe that  the tree level contribution may attain  $115$ GeV for $\tan(\beta)=2$ which is in a good agreement  with the predictions  of the  Eq. (\ref{ma}). In other words, our results confirm that  Eq. (\ref{ma})  is a valid approximation for the tree level contribution to the Higgs-like  boson mass in the minimal SUSY331 model. Thus, the robustness of the Higgs-like boson mass at tree level is   dictated by $\tan(\beta)$ such that  the smaller $\tan(\beta)$, the larger tree level Higgs-like boson mass. 
 
 Nevertheless, we know that loop correction to the Higgs mass are considerable, enforcing us to conclude that   $\tan(\beta) \geq 2$ and that the model may support a stop with mass below the TeV  scale as required by naturalness principle.

For sake of completeness, we obtain the eigenvector  associated to the eigenvalue $125$ GeV  for the following set of values  of the parameters
\bea 
&&\tan (\beta )=2.7, \,\,\,\tan (\beta_\eta)=10,\,\,\,\tan (\beta_\rho)=10,\,\,\,\tan (\beta_\chi)=1.2,\,\,\,v_{\eta }=22.3\,\mathrm{GeV},\,\,\,v_{\eta^{\prime}}=223.2\,\mathrm{GeV},\nonumber \\
&&v_{\rho }=8.1\, \mathrm{GeV},\,\,\,v_{\rho^{\prime}}=81\,\mathrm{GeV},\,\,\,v_{\chi }=1763.6 \,\mathrm{GeV},\,\,\,v_{\chi^{\prime}}=2116.4\,\mathrm{GeV}, \,\,\, f_1=0.01 ,\,\,\,f_1^{\prime}=0.6 ,\nonumber \\
&& k_1=-500 \,\mathrm{GeV},\,\,\,k_2=-600\, \mathrm{GeV},\,\,\,b_{\eta}=-55000 \,\mathrm{GeV}^2,\,\,\,b_{\rho }=-1.7\times 10^6\,\mathrm{GeV}^2,\nonumber \\
&&b_{\chi }=-1.4\times 10^6 \,\mathrm{GeV}^2, \mu _{\rho }=-1700 \,\mathrm{GeV},\,\,\, \mu _{\eta }=-550 \,\mathrm{GeV},\,\,\, \mu _{\chi }=-1400 \,\mathrm{GeV}.
\label{peigenvector}
\eea
In this case, the  eigenvector corresponding to the 125 GeV Higgs is 
\be
h = -0.93R_{\eta^{\prime}} -0.34R_{\rho^{\prime}} -0.09R_\eta -0.03R_\rho -0.05R_{\chi^{\prime}} -0.06R_\chi.
\label{higgseigenvector}
\ee
See that the eigenvector $h$  is  composed mainly by  $R_{\eta^{\prime}}$ and $R_{\rho^{\prime}}$.  For any other choice  of the  set of values of the parameters, the eigenvector will keep being dominantly a composition of $R_{\eta^{\prime}}$ and $R_{\rho^{\prime}}$.
\subsection{One Loop Level Contribution}
In the MSSM a 125 GeV Higgs  demands robust loop corrections which requires  stop heavy enough such that threaten the naturalness principle. In the minimal SUSY331 model things are a little different once tree level contribution to the Higgs-like boson mass may surpass $100$ GeV.  

In what follows we calculate numerically, by  employing  the package Spheno,  the Higgs-like boson mass but taking into account one loop  corrections. Our results are displayed  in  graphics showing the behavior of our Higgs-like boson mass with the parameters  $M_s$, $X_t$ and $v_{331}$. For the other parameters, we take the values
\begin{eqnarray}
\centering
f_1 = 0.001 \,&,& f_1^{\prime} = 0.6\,,  \nonumber \\
k_1 =-500\, \mathrm{GeV} \,&,&  k_2 = -600\, \mathrm{GeV}\,,\nonumber \\
\mu_{\eta} = -550\, \mathrm{GeV}, \, \mu_{\rho} = &-&1700\,\mathrm{GeV}\,,\mu_{\chi} = -1400\,\mathrm{GeV}\,, \nonumber \\ 
\tan{\beta} = 2, \tan{\beta_\eta}=&& \tan{\beta_\rho} = 10, \tan{\beta_\chi} = 1.20\,,  \nonumber \\
A_0, B_0 &=& 1000\, \mathrm{GeV}.
\label{fixedoarameters}
\end{eqnarray}

\begin{figure}[h!]
 \includegraphics[width=8cm,height=6cm]{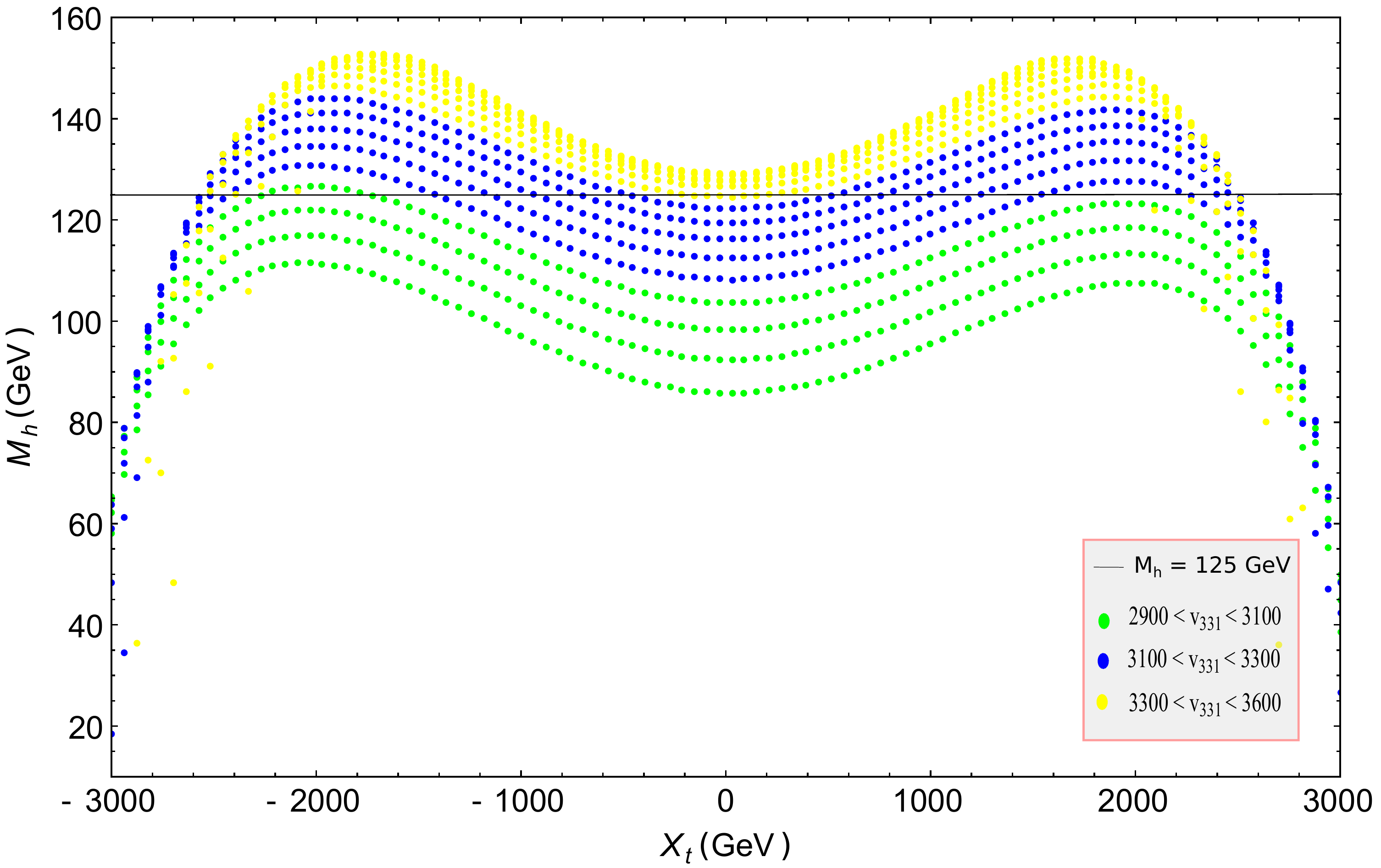}
  \caption{$M_h$ vs $X_t$ for various values of $v_{331}$. We took  $m_{SUSY_L}=m_{SUSY_R}=[500,1000]$ GeV.}
  \label{higgsXt}
\end{figure}

In FIG. \ref{higgsXt} we plot $M_h$ versus $X_t$.  As we can see in that plot, small $X_t$ is allowed for large value of  $v_{331}$, while large $X_t$ is allowed by small values of  $v_{331}$. Moreover, observe that we may obtain a 125 GeV Higgs in the minimal SUSY331 model even for $X_t=0$. We highlight this result because it is not allowed in the MSSM.
%%%%%%%%%
\begin{figure}[h!]
\includegraphics[width=8cm,height=6cm]{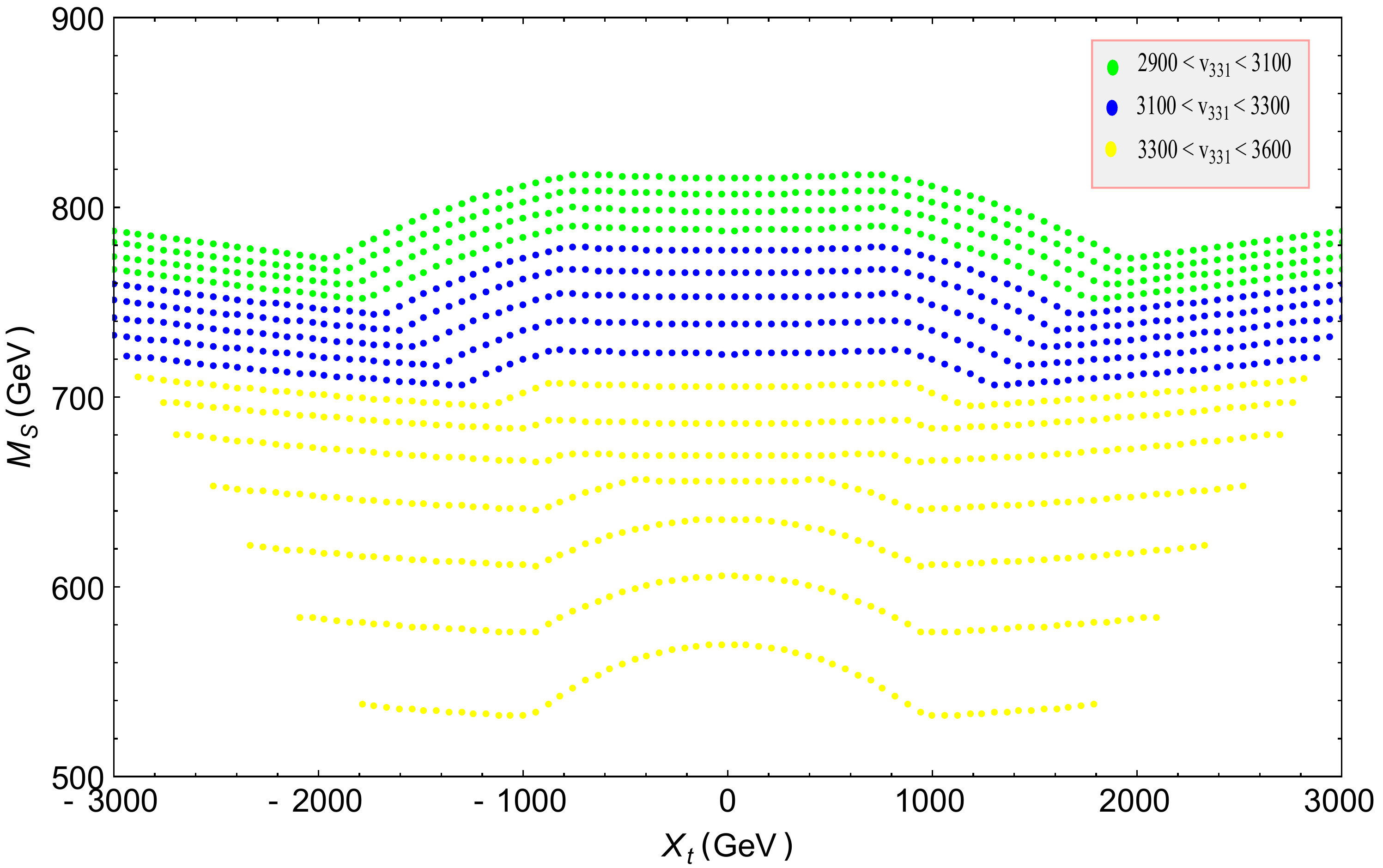}
  \caption{$M_s$ versus $X_t$ for various values of $v_{331}$. }
  \label{XtMs}
\end{figure}

\begin{figure}[h!]
 \includegraphics[width=8cm,height=6cm]{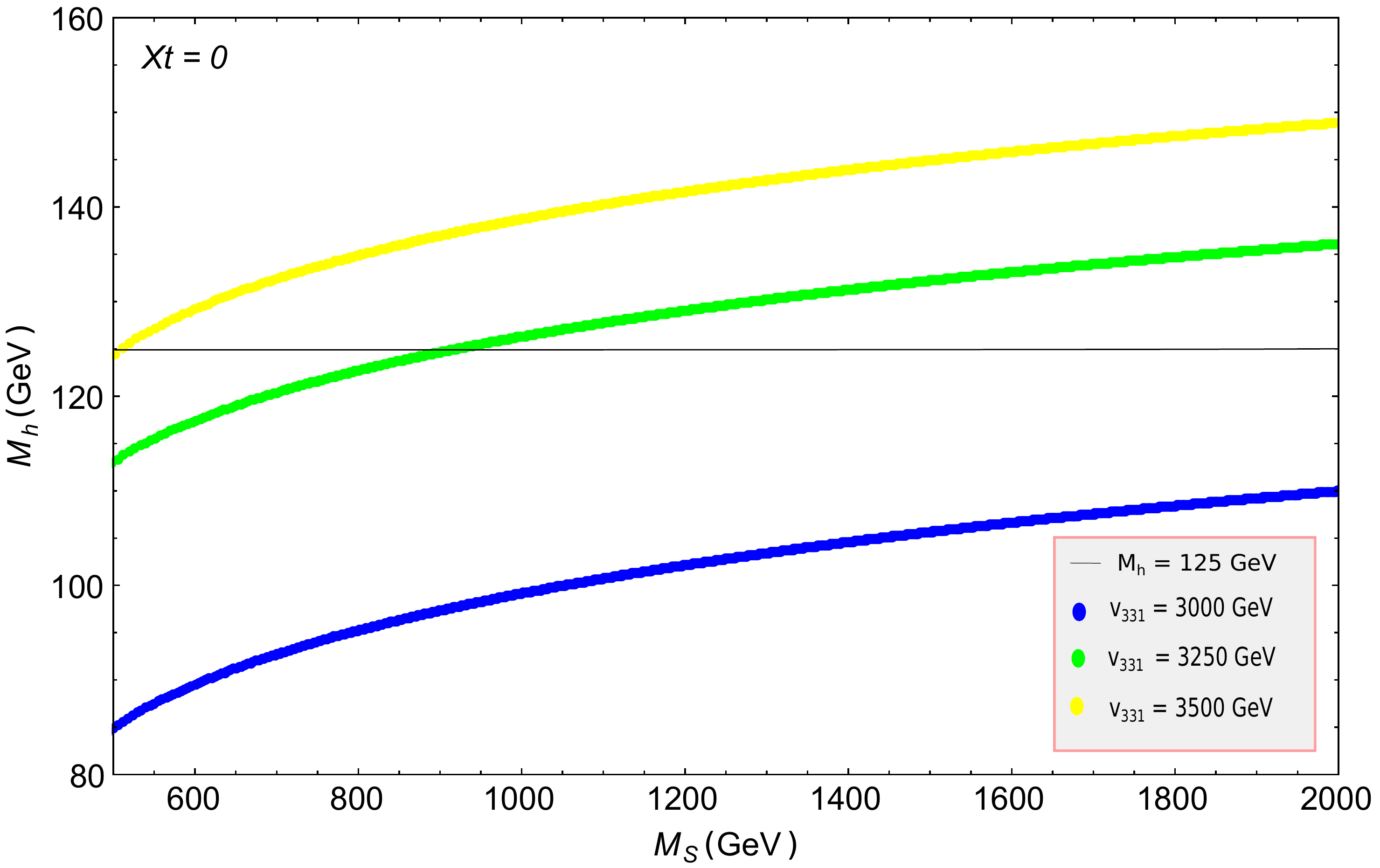}
  \caption{$M_h$ versus $M_s$ for $X_t=0$ for three different values of $v_{331}$.  }
  \label{Xtmsoft}
\end{figure}

\begin{figure}[h!]
 \includegraphics[width=11cm,height=12cm]{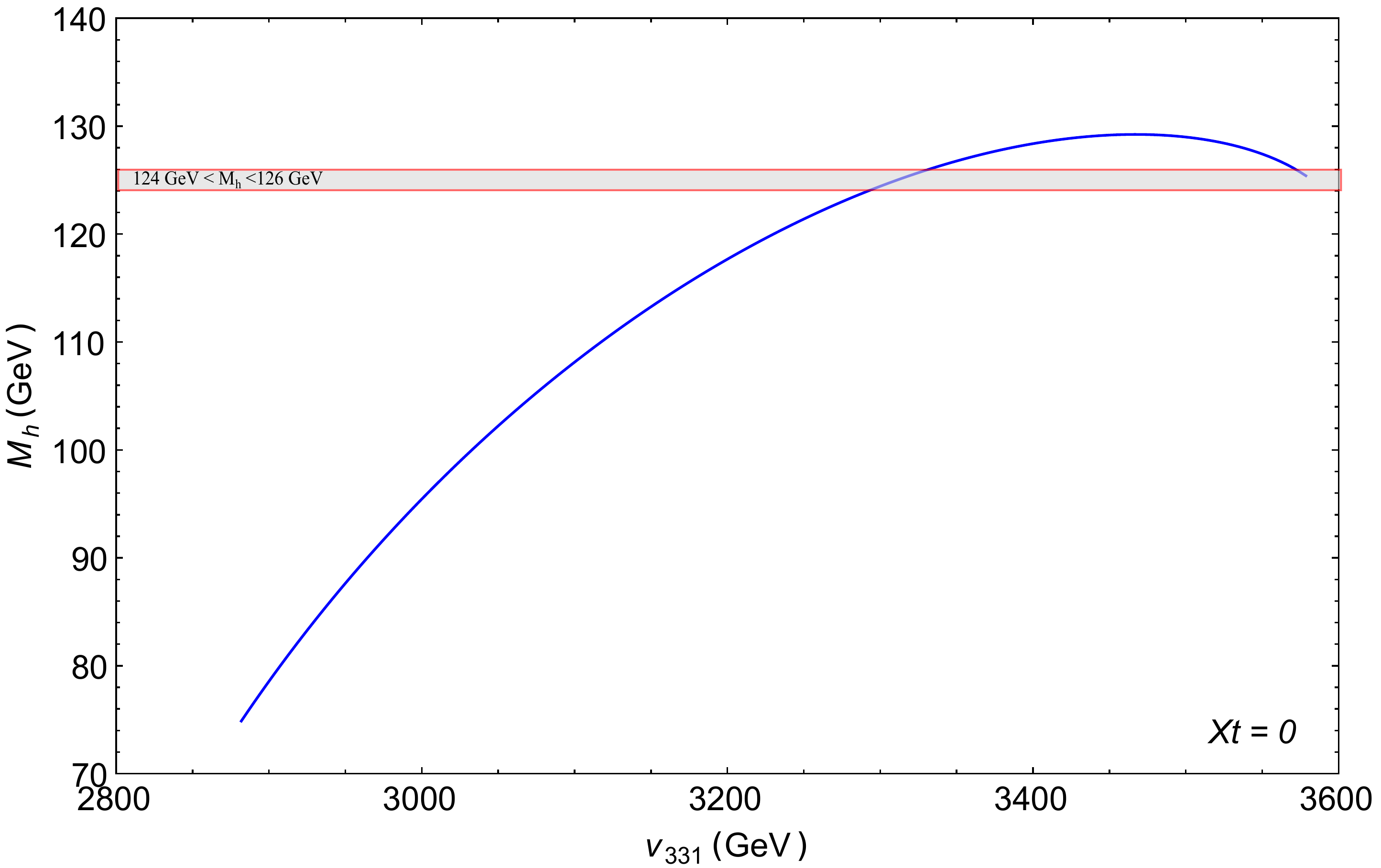}
  \caption{$M_h$ versus $v_{331}$ for $X_t=0$.}
  \label{Mhv331}
\end{figure}

We also obtained the $2 \times 2$ mass matrix for the stops and diagonalized it with the Spheno package. The result is displayed in FIG. \ref{XtMs}  where we show the behavior of  $M_s$  with $X_t$. Note that the larger $v_{331}$, the smaller $M_s$. In  FIG. \ref{Xtmsoft} we show the behavior of $M_h$ with $M_s$ for the specific  case $X_t=0$, and in FIG. \ref{Mhv331} we show the behavior of $M_h$ with $v_{331}$ for $X_t=0$. Perceive that the model provides easily  a $125$ GeV
Higgs boson for  $M_s$ and $X_t$, both, below the TeV scale. 

As we can see, the minimal SUSY331 model naturally recovers the standard Higgs boson mass, but  we know that it is  different from the MSSM  one because of the peculiarities of the minimal SUSY331 model. Thus, in order to conclude definitely that our Higgs-like boson recovers all the observed properties of  the standard Higgs,  we need to calculate its branching ratios and the respective signal strengths and confront the results with the experimental data  measured by LHC.

\section{Branching Ratios and signal strengths }
\label{sec3}
\begin{figure}[h!]
 \includegraphics[width=8cm,height=6cm]{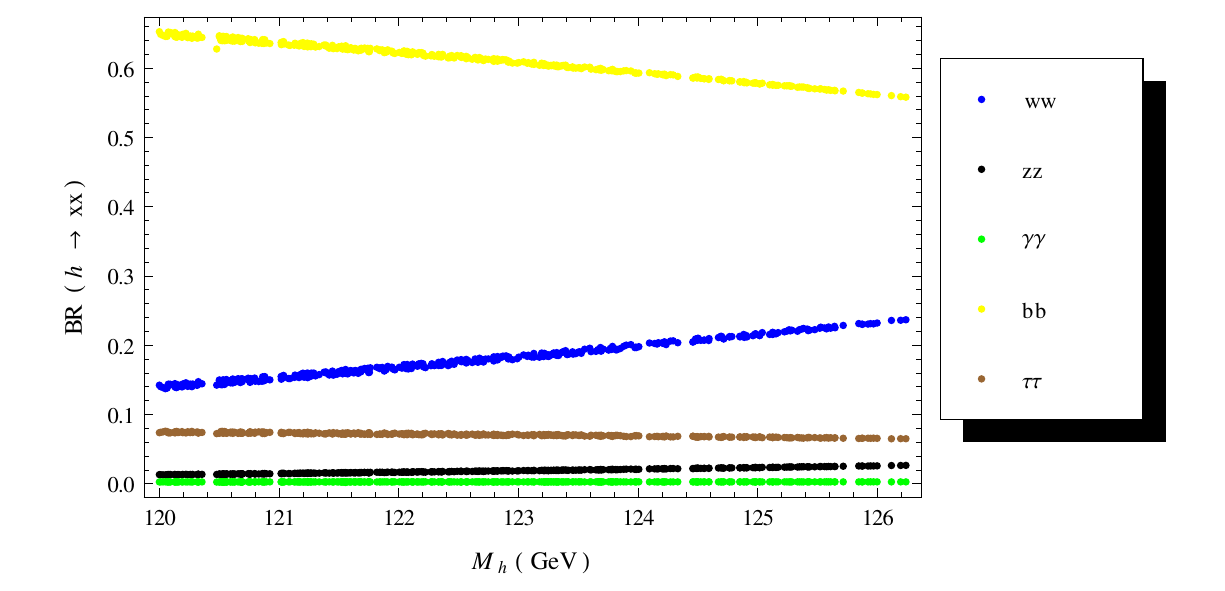}
  \includegraphics[width=8cm,height=6cm]{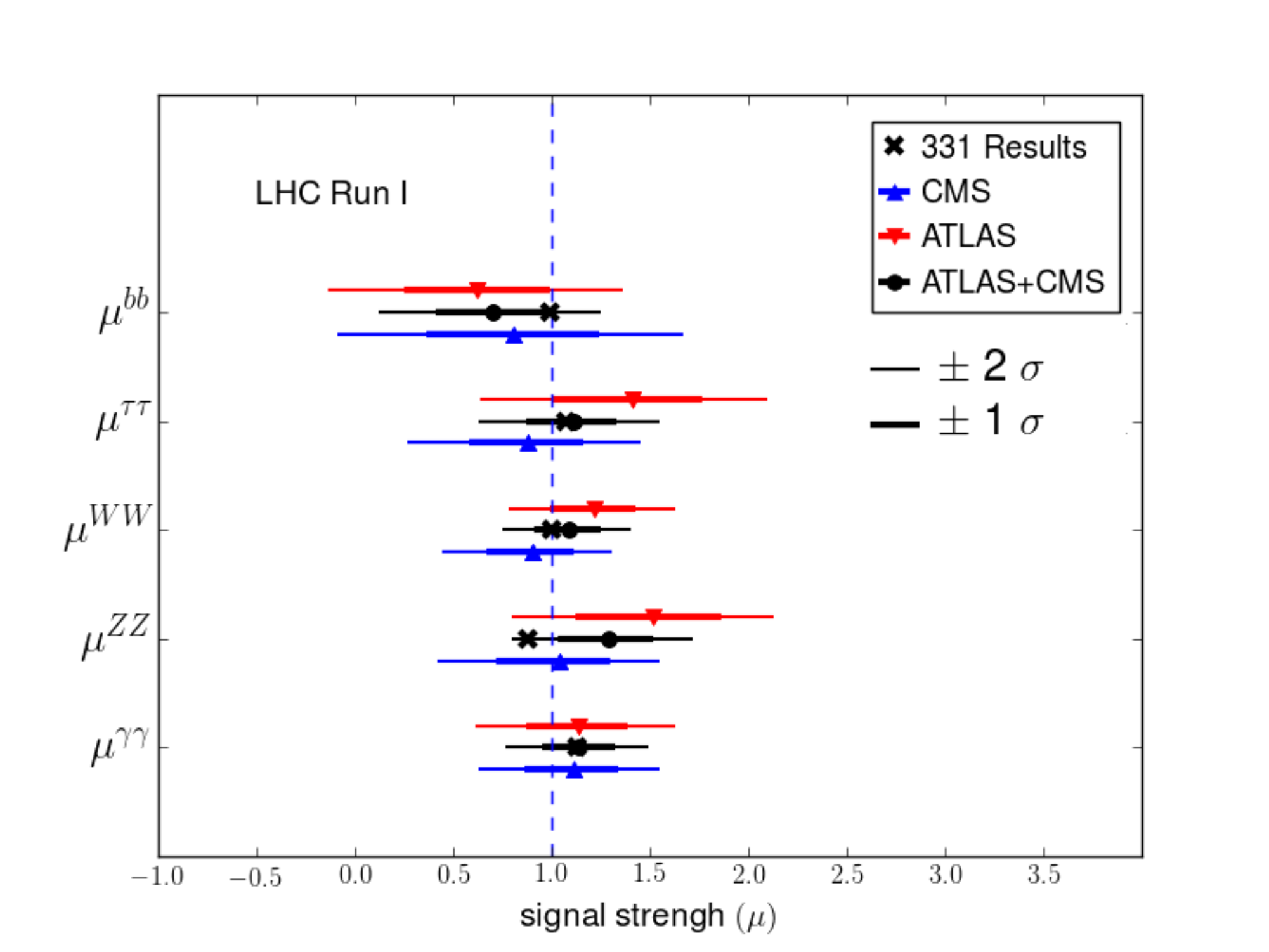}
  \caption{Left: The dominant branching ratios for our Higgs-like boson as function of its mass. Right: Decay  signal strengths taking into account the combination of ATLAS and CMS data. The error bars indicate the $1\sigma$ (thick lines) and $2\sigma$ (thin lines) intervals. The combined results show a remarkable agreement with the SM prediction (normalized to $\mu = 1$)}
 \label{BR}
\end{figure}

In what concerns the standard Higgs,   any extension of the SM must possess a scalar with 125 GeV of mass and couplings with the standard particles that fit the measured branching ratios and signal strengths \cite{Heinemeyer:2013tqa}. This is the reason why we wonder if the  Higgs-like boson discussed here recovers the  Higgs   branching ratios and signal strengths as observed by LHC. Our analysis is done for  the following set of  parameters 
\begin{eqnarray}
\centering
&&f_1^{\prime} = 0.6\, , \, -0.03 \leq  f_1 \leq 0.2\,, \nonumber \\
&&k_1 = -500\, \mathrm{GeV} \,,  k_2 = -600 \, \mathrm{GeV}\,,\nonumber \\
&&\mu_{\eta} = -550 \, \mathrm{GeV}, \, \mu_{\rho} = -1700\,\mathrm{GeV}\,,\mu_{\chi} = -1400\,\mathrm{GeV}\,, \nonumber \\ 
&& \tan{\beta_\eta}, \,\, \tan{\beta_\rho} = 10 , \, \tan{\beta_\chi} = 1.20\,,   \nonumber \\
&&1000 \, \mathrm{GeV} \leq v_{331} \leq 5000 \, \mathrm{GeV} \, , \, 2 \leq \,  \tan{\beta} \, \leq 10 , \nonumber \\
&&500 \, \mathrm{GeV} \leq m_{SUSY_L}, \, m_{SUSY_R} \leq 2000 \, \mathrm{GeV}, \nonumber \\
&&X_t=0,\,\,\,A_0, B_0 = 1000 \, \mathrm{GeV}.
\label{scan3}
\end{eqnarray}
As we know, the  Higgs prefers to decay  into pairs of $ b \bar  b$,  $W W^*$, $\tau \bar \tau$, $Z Z^*$ and $\gamma \gamma$. We restrict our investigation for these channels.  Our results are shown in  FIG. \ref{BR}. Perceive that the  predictions are in perfect agreement with the values measured by ATLAS\cite{Aad:2015gba} and CMS\cite{CMS:2014ega}.  In this point of the work, we are secure in affirming that the minimal SUSY331 model is privileged in what concerns Higgs physics since it posseses a Higgs-like boson with mass and couplings with the standard particles that recovers its observed properties and respecting the naturalness principle.
%%%%%%%%%%%%%%%%%%%
%%%%%%%%
\section{Flavor Changing Neutral Current as Signature of our Higgs-like boson}
\label{sec4}
\begin{figure}[h!]
 \includegraphics[width=8cm,height=6cm]{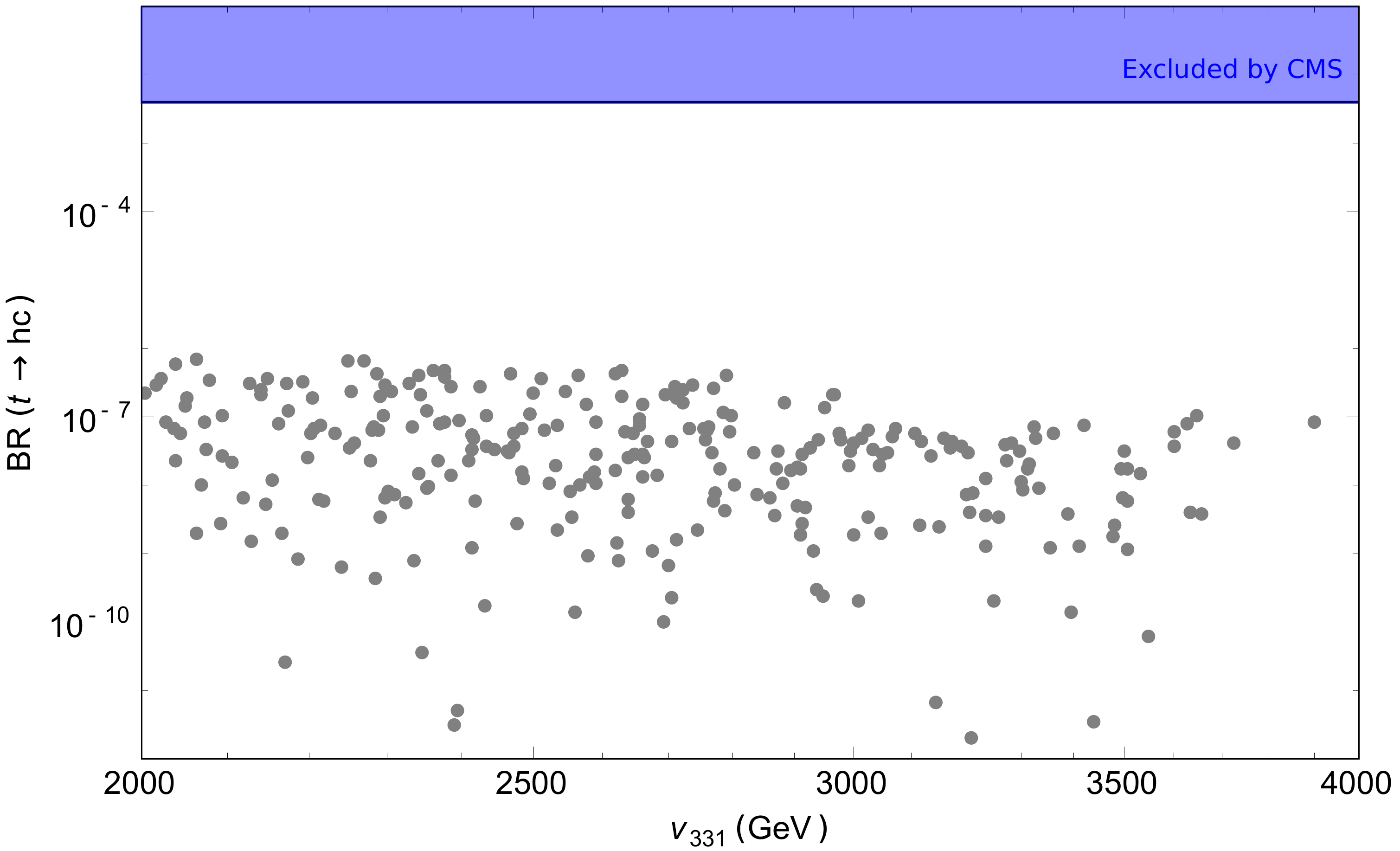}
  \includegraphics[width=8cm,height=6cm]{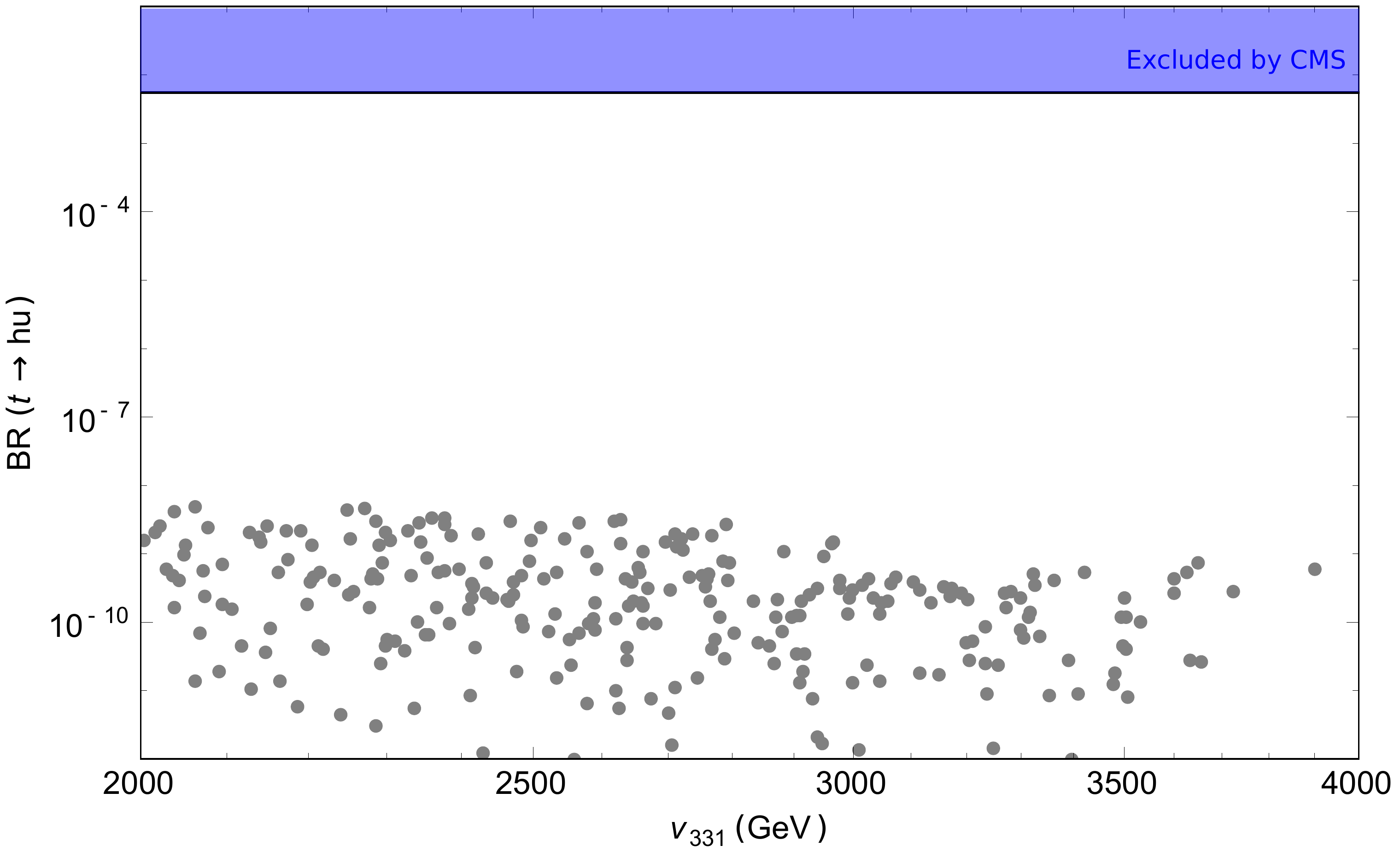}
  \caption{Calculation of the branching ratios  for the top decays $t\rightarrow hc$ and $t\rightarrow hu$. }
 \label{FCTD}
\end{figure}

%%%%%%%%%%%%%%%%%
As we saw above, the lightest Higgs of the minimal SUSY331 model fulfills the conditions to be a SM-like Higgs. However, it is important to recall that our Higgs-like bosom is different from the SM-like one in many aspects  with the major one being that itintermediates flavor changing processes involving quarks already at tree level.   This is a consequence of  quark masses having more than one source, as depicted in the superpotential of the model. According to the last line of the superpotential, and the fact that our Higgs-like boson is mostly a composition of $R_{\eta^{\prime}}$ and $R_{\rho^{\prime}}$,  the dominant flavor changing processes are those  involving the third family of quarks. We studied all the possible processes here and obtained that the decays $h \rightarrow bs\,,\,bd$ are very surppressed even in relation to the loop predictions of the SM\cite{Benitez-Guzman:2015ana}.   Thus, the significant processes are those  involving the Yukawa interactions $t-h-u$ and $t-h-c$. For this case, the signature of our Higgs-like boson is through top quark decays via Higgs-mediated flavor-changing processes \cite{Khachatryan:2016atv}. The behavior of the branching ratio of these decays as function of $v_{331}$ is presented in FIG. \ref{FCTD}. Our calculations were done for the following region of the parameter space
%%%
\begin{eqnarray}
\centering
&&f_1^{\prime} = 0.6\, , \, -0.03 \leq  f_1 \leq 0.2\,, \nonumber \\
&&k_1 = -500\, \mathrm{GeV} \,,  k_2 = -600 \,\mathrm{GeV}\,,\nonumber \\
&&\mu_{\eta} = -550 \, \mathrm{GeV}, \, \mu_{\rho} = -1700\,\mathrm{GeV}\,,\mu_{\chi} = -1400\,\mathrm{GeV}\,, \nonumber \\ 
&& \tan{\beta_\eta}, \,\, \tan{\beta_\rho} = 10 , \, \tan{\beta_\chi} = 1.20\,,   \nonumber \\
&&2000 \, \mathrm{GeV} \leq v_{331} \leq 4000 \, \mathrm{GeV} \, , \, 2 \leq \,  \tan{\beta} \, \leq 4 , \nonumber \\
&&800 \, \mathrm{GeV} \leq m_{SUSY_L}, \, m_{SUSY_R} \leq 1200 \, \mathrm{GeV}, \nonumber \\
&&-5000 \leq X_t \leq 5000 ,\,\,\,A_0, B_0 = 1000 \, \mathrm{GeV}.
\label{scan3}
\end{eqnarray}
Although our results are far below the excluded region, we hope this to be probed in the next generation of collider. 

%%%%%%
%%%%%%%%

\section{Stability of the potential}
\label{sec5}
\begin{figure}[h!]
 \includegraphics[width=8cm,height=6cm]{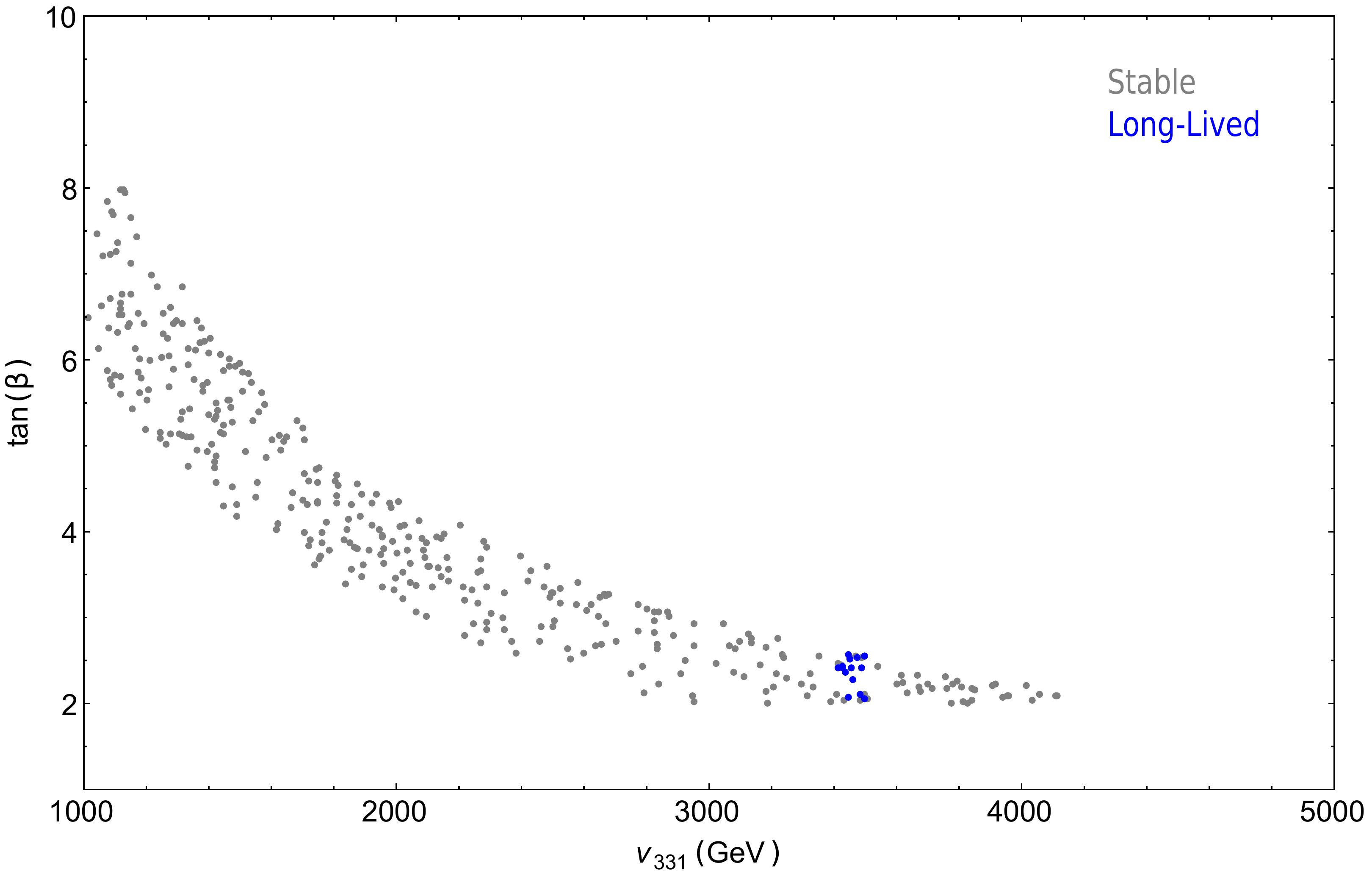}
 \caption{Points  in the plane $\tan(\beta)$ versus $v_{331}$ that guarantee stability of the potential.}
 \label{vev}
\end{figure}

%%%%%%%%%%%%%%%%%%%%%
In this section we check if  that  region of parameter space considered throughout this paper is so that the potential of the model is stable for such value of the parameters.  In practical terms, we have to be sure that the minimum of the potential we are considering is in fact  the global one. We check this by means of the Vevacious Package \cite{Camargo-Molina:2014pwa,Camargo-Molina:2013qva}. We export the model implemented in the Sarah package to the Vevacious one and scanned that region of the parameter space used in the results above. Our result is presented  in the plane  $\tan(\beta)$ versus $v_{331}$ displayed in FIG. \ref{vev} .
\begin{eqnarray}
\centering
&&f_1^{\prime} = 0.6\, , \, -0.03 \leq  f_1 \leq 0.2\,, \nonumber \\
&&k_1 = -500\, \mathrm{GeV} \,,  k_2 = -600\, \mathrm{GeV}\,,\nonumber \\
&&\mu_{\eta} = -550 \, \mathrm{GeV}, \, \mu_{\rho} = -1700\,\mathrm{GeV}\,,\mu_{\chi} = -1400\,\mathrm{GeV}\,, \nonumber \\ 
&& \tan{\beta_\eta}, \,\, \tan{\beta_\rho} = 10 , \, \tan{\beta_\chi} = 1.20\,,   \nonumber \\
&&1000 \, \mathrm{GeV} \leq v_{331} \leq 5000 \, \mathrm{GeV} \, , \, 2 \leq \,  \tan{\beta} \, \leq 10 , \nonumber \\
&&500 \, \mathrm{GeV} \leq m_{SUSY_L}, \, m_{SUSY_R} \leq 2000 \, \mathrm{GeV}, \nonumber \\
&&-3000\leq X_t \leq3000,\,\,\,A_0, B_0 = 1000 \, \mathrm{GeV}.
\label{scan3}
\end{eqnarray}
Perceive that, except for  a small region of the parameter space presenting long lived behavior, but with  decay time larger than the age of our Universe (blue region), major part of the region of the parameter space considered in this work   is associated to a stable minimum of the  potential. 
%%%%%%%%%%%%%%%%%%

%%%%%%
\section{Conclusions}
\label{sec6}
In this work we developed the Higgs sector of the minimal SUSY331 model with the focus on the lightest Higgs. We obtained  its mass and couplings with the standard particles.  With regard to its mass, tree level contribution may surpass the usual MSSM prediction and attain more than 100 GeV. Consequently, a 125 GeV Higgs will demand  feeble  loop corrections. In our calculations we showed that a 125 GeV Higgs is compatible with stop with mass below the TeV scale running in the loops. This is a remarkable result  concerning the  naturalness principle. 

Although Higgs mass is the most important aspect of the Higgs physics, a complete work demands we extend our investigation to the Higgs couplings with the standard particles. We performed such an analysis  and confirmed that they recover the observed pattern of branching ratios and signal strengths for its dominant processes. Finally we discussed the signature of our Higgs-like boson which manifests in the form of flavor changing processes intermediated by the Higgs. The most relevant ones  are the top decay channels  $t\rightarrow h+u$ and $t\rightarrow h+c$. We studied the behavior of these processes and concluded that they are out of reach of the LHC but, perhaps, may be probed in future colliders. We also checked if the region of parameter space considered in this work is compatible with the stability of the potential.  We obtained that, except for  a small region of the parameter space presenting long lived behavior, but with  decay time larger than the age of our Universe, major part of it  belongs to those points where the potential has a stable global minimum. All of this leads us to conclude that the minimal SUSY331 model is an interesting supersymmetric model for particle physics. 
%%%%%%%%%%%%%%%%%%%%%%%%%%%%%%%%%%%%%%%%%%%%%%%%%%%%%

\acknowledgments
We thank Felipe Ferreira for useful discussions. This work was supported by Conselho Nacional de Pesquisa e
Desenvolvimento Cient\'{i}fico- CNPq (C.A.S.P,  P.S.R.S.). CS is supported by CAPES/PDSE Process 88881.134759/2016-01.
%%%%%%%%%%%%%%%%%%%%%%%%%%%%%%%%%%%%%%%%%%%%%%
%%%%%%%%
%%%%%%%%%%%%%%%
\bibliography{S331min.bib}
%%%%%%
\appendix
\section*{Appendix A}
\begin{eqnarray}
\left\langle \frac{\partial V}{\partial \rho^{0}}\right\rangle_0 &=& (m^{2}_\rho + \mu_\rho^2) v_{\rho} + b_{\rho} v_{\rho^\prime} +\frac{1}{12} g^2 v_{\rho} \left(-v_{\eta}^2  + v_{\eta^\prime}^{2} - v_{\chi}^2  + v_{\chi^\prime}^{2} + 2 v_{\rho}^2 - 2 v_{\rho^\prime}^{2} \right) \nonumber \\
&+& \frac{1}{2} g_N^2 v_{\rho} \left( -v_{\chi}^2  + v_{\chi^\prime}^{2} + v_{\rho}^2 - v_{\rho^\prime}^{2} \right) +\frac{1}{\sqrt{2}} f_1^\prime \left(  v_{\eta^\prime} v_{\chi^\prime} \mu_{\rho} \right)  \nonumber \\
&+& \frac{1}{\sqrt{2}} f_1 \left( \frac{1}{\sqrt{2}} v_\eta^{2} v_{\rho} + \frac{1}{\sqrt{2}} v_{\chi}^{2} v_{\rho} + v_{\eta} v_{\chi^\prime} \mu_{\chi} + v_{\eta^\prime} v_\chi \mu_{\eta} \right)  + \frac{1}{\sqrt{2}} k_1 v_\eta v_{\chi} = 0 \nonumber ,\\
\left\langle \frac{\partial V}{\partial \rho^{\prime 0}}\right\rangle_0 &=& (m^{2}_{\rho^{\prime}} + \mu_{\rho}^2) v_{\rho^\prime} + b_{\rho} v_{\rho} +\frac{1}{12} g^2 v_{\rho^\prime} \left(v_{\eta}^2  - v_{\eta^\prime}^{2} + v_{\chi}^2  - v_{\chi^\prime}^{2} - 2 v_{\rho}^2 + 2 v_{\rho^\prime}^{2} \right) \nonumber \\
&+& \frac{1}{2} g_N^2 v_{\rho^\prime} \left( -v_{\chi}^2  + v_{\chi^\prime}^{2} + v_{\rho}^2 - v_{\rho^\prime}^{2} \right) +\frac{1}{\sqrt{2}} f_1 \left(  v_{\eta} v_{\chi} \mu_{\rho} \right)  \nonumber \\
&+& \frac{1}{\sqrt{2}} f_1^\prime \left( \frac{1}{\sqrt{2}} v_{\eta^\prime}^{2} v_{\rho^\prime} + \frac{1}{\sqrt{2}} v_{\chi^\prime}^{2} v_{\rho^\prime} + v_{\eta} v_{\chi^\prime} \mu_{\eta} + v_{\eta^\prime} v_\chi \mu_{\chi} \right)  + \frac{1}{\sqrt{2}} k_2 v_{\eta^\prime} v_{\chi^\prime} = 0 \nonumber ,\\
\left\langle \frac{\partial V}{\partial \eta^{0}}\right\rangle_0 &=& (m^{2}_\eta + \mu_\eta^2) v_{\eta} + b_{\eta} v_{\eta^\prime} +\frac{1}{12} g^2 v_{\eta} \left(-v_{\rho}^2  + v_{\rho^\prime}^{2} - v_{\chi}^2  + v_{\chi^\prime}^{2} + 2 v_{\eta}^2 - 2 v_{\eta^\prime}^{2} \right) \nonumber \\
&+& \frac{1}{2} f_1^\prime \left(  v_{\rho^\prime} v_{\chi^\prime} \mu_{\eta} \right)  + \frac{1}{\sqrt{2}} f_1 \left( \frac{1}{\sqrt{2}} v_\rho^{2} v_{\eta} + \frac{1}{\sqrt{2}} v_{\chi}^{2} v_{\eta} + v_{\rho} v_{\chi^\prime} \mu_{\chi} + v_{\rho^\prime} v_\chi \mu_{\rho} \right)  \nonumber \\
&+& \frac{1}{\sqrt{2}} k_1 v_\rho v_{\chi} = 0 \nonumber ,\\
\left\langle \frac{\partial V}{\partial \eta^{\prime 0}}\right\rangle_0 &=& (m^{2}_{\eta^{\prime}} + \mu_{\eta}^2) v_{\eta^\prime} + b_{\eta} v_{\eta} +\frac{1}{12} g^2 v_{\eta^\prime} \left(v_{\rho}^2  - v_{\rho^\prime}^{2} + v_{\chi}^2  - v_{\chi^\prime}^{2} - 2 v_{\eta}^2 + 2 v_{\eta^\prime}^{2} \right) \nonumber \\
&+& \frac{1}{\sqrt{2}} f_1 \left(  v_{\rho} v_{\chi} \mu_{\eta} \right)  + \frac{1}{\sqrt{2}} f_1^\prime \left( \frac{1}{\sqrt{2}} v_{\rho^\prime}^{2} v_{\eta^\prime} + \frac{1}{\sqrt{2}} v_{\chi^\prime}^{2} v_{\eta^\prime} + v_{\rho} v_{\chi^\prime} \mu_{\rho} + v_{\rho^\prime} v_\chi \mu_{\chi} \right) \nonumber \\
& +& \frac{1}{\sqrt{2}} k_2 v_{\rho^\prime} v_{\chi^\prime} = 0 \nonumber , \\
\left\langle \frac{\partial V}{\partial \chi^{0}}\right\rangle_0 &=& (m^{2}_\chi + \mu_\chi^2) v_{\chi} + b_{\chi} v_{\chi^\prime} +\frac{1}{12} g^2 v_{\chi} \left(-v_{\eta}^2  + v_{\eta^\prime}^{2} - v_{\rho}^2  + v_{\rho^\prime}^{2} + 2 v_{\chi}^2 - 2 v_{\chi^\prime}^{2} \right) \nonumber \\
&+& \frac{1}{2} g_N^2 v_{\chi} \left(v_{\chi}^2  - v_{\chi^\prime}^{2} - v_{\rho}^2 + v_{\rho^\prime}^{2} \right) +\frac{1}{\sqrt{2}} f_1^\prime \left(  v_{\eta^\prime} v_{\rho^\prime} \mu_{\chi} \right)  \nonumber \\
&+& \frac{1}{\sqrt{2}} f_1 \left( \frac{1}{\sqrt{2}} v_\eta^{2} v_{\chi} + \frac{1}{\sqrt{2}} v_{\rho}^{2} v_{\chi} + v_{\eta} v_{\rho^\prime} \mu_{\rho} + v_{\eta^\prime} v_\rho \mu_{\eta} \right)  + \frac{1}{\sqrt{2}} k_1 v_\eta v_{\rho} = 0 \nonumber , \\
\left\langle \frac{\partial V}{\partial \chi^{\prime 0}}\right\rangle_0 &=& (m^{2}_{\chi^{\prime}} + \mu_{\chi^\prime}^2) v_{\chi^\prime} + b_{\chi} v_{\chi} +\frac{1}{12} g^2 v_{\chi^\prime} \left(v_{\eta}^2  - v_{\eta^\prime}^{2} + v_{\rho}^2  - v_{\rho^\prime}^{2} - 2 v_{\chi}^2 + 2 v_{\chi^\prime}^{2} \right) \nonumber \\
&+& \frac{1}{2} g_N^2 v_{\chi^\prime} \left( -v_{\chi}^2  + v_{\chi^\prime}^{2} + v_{\rho}^2 - v_{\rho^\prime}^{2} \right) +\frac{1}{\sqrt{2}} f_1 \left(  v_{\eta} v_{\rho} \mu_{\chi} \right)  \nonumber \\
&+& \frac{1}{\sqrt{2}} f_1^\prime \left( \frac{1}{\sqrt{2}} v_{\eta^\prime}^{2} v_{\chi^\prime} + \frac{1}{\sqrt{2}} v_{\rho^\prime}^{2} v_{\chi^\prime} + v_{\eta} v_{\rho^\prime} \mu_{\eta} + v_{\eta^\prime} v_\rho \mu_{\rho} \right)  + \frac{1}{\sqrt{2}} k_2 v_{\eta^\prime} v_{\rho^\prime} = 0 .
\label{minimumconditions}
\end{eqnarray}

\end{document}